%% file: main.tex
\documentclass[letterpaper, 10 pt, conference]{ieeeconf}  

\IEEEoverridecommandlockouts                              

\usepackage{amsmath,amsfonts,bm,bbm,ifthen,mathtools}
\usepackage{amssymb}

\usepackage{tikz}
\usetikzlibrary{shapes,arrows,positioning,fit}

\usepackage{subcaption}
\usepackage{orcidlink}
\usepackage{hyperref}

\newlength{\tikzw}
\title{\LARGE \bf
Interpreted Higher-Dimensional Automata\\
for\\
Concurrent Discrete-Event Control
}

\author{Dylan Bellier$^{2}$\orcidlink{0000-0003-4763-5655}, Gregory Faraut$^{1}$\orcidlink{0000-0003-0388-932X} , Yan Monier$^{1}$\orcidlink{0000-0002-7683-2159}, Philipp Schlehuber-Caissier$^{2}$\orcidlink{0000-0002-6611-9659}
\thanks{$^{1}$Université Paris-Saclay, ENS Paris-Saclay, LURPA, France.
        {\tt\small firstname.lastname@ens-paris-saclay.fr}}%
\thanks{$^{2}$SAMOVAR, Télécom SudParis, Institut Polytechnique de Paris, France.
Partially funded by the Academic and Research Chair ``Architecture des Systèmes Complexes''
Dassault Aviation, Naval Group, Dassault Systèmes, KNDS France, Agence de l'Innovation de Défense, 
Institut Polytechnique de Paris}%
}

\input{macros.tex}

\begin{document}


\maketitle
\thispagestyle{empty}
\pagestyle{empty}

\begin{abstract}

In recent years the theory of Higher Dimensional Automata (HDA) has seen significant 
advances from a theoretical point of view, reflecting standard automata theory 
(for instance \cite{Hdalang}). 
There have also been first attempts to use the mathematical framework provided by 
HDAs to known problems, in particular Petri Net analysis (see 
\cite{DBLP:journals/tcs/Glabbeek06, DBLP:conf/apn/AmraneBFHS25}).
However \textit{real-world} applications are still lacking and issues from 
\textit{real-world} system, as concurrency, is still opened in the context of controller generation.
In this work we show how the framework of HDAs can be adapted to help transforming
controllers given as interpreted Petri nets (IPN) into an actual closed loop controller 
and how the HDA helps in identifying ambiguous or even contradictory specifications that 
remain ``hidden'' in the IPN. 
We demonstrate the feasibility by connecting the obtain controller to a virtual 
environment for closed loop control, exemplified by an industrial example.

\end{abstract}

\section{Introduction}
\label{sec:Intro}

\input{intro.tex}

\section{Interpreted Petri Nets}
\label{sec:IPN}

\input{pndef.tex}

\section{Higher Dimensional Automata}
\label{sec:HDA}

\input{hdadef.tex}

\section{Introducing Interpreted HDAs}
\label{sec:IHDA}

\input{ihdadefPSC.tex}

\section{Manufacturing Example}
\label{sec:ME}

\input{example.tex}

\subsection{Error Detection and Correction}
\label{sec:ME_TEDC}

\input{error_detection.tex}

\subsection{IHDA controller in simulation tool}
\label{sec:simul}
\input{simulation.tex}


\section{CONCLUSIONS}
\label{sec:Conc}

\input{conclusion}



\bibliographystyle{plain}
\bibliography{mybib}

\end{document}

%% file: macros.tex
\usepackage{xspace}
\usepackage{leftindex}

\newtheorem{definition}{Definition}

\newtheorem{lemma}{Lemma}

\usepackage{tikz}
\usetikzlibrary{arrows, automata, shapes, calc, petri, arrows.meta}

\tikzset{->, auto, >=stealth', font=\small}
\tikzset{state/.style={shape=circle, draw, fill=white, initial text=, inner sep=.5mm, minimum size=1.5mm}}
\tikzset{accepting/.style=accepting by arrow}
\tikzset{state with output/.style={shape=rectangle split, rectangle split parts=2, draw, fill=white, initial text=, inner sep=1mm}}
\tikzset{place/.style={shape=circle, draw, minimum size=4mm}}
\tikzset{transition/.style={fill,minimum width=5mm,minimum height=1mm, inner sep=.1mm}}

\newcommand{\textfun}[1]{\textsf{#1}}

\newcommand{\ie}{i.e.\xspace}

\newcommand{\Alphabet}{\Sigma}

\newcommand{\petrinet}[1][]{N_{#1}}
\newcommand{\petrinetDef}[1][]{\tuple{\Places[#1], \Transitions[#1], \FlowRelation[#1]}}

\newcommand{\Places}[1][]{S_{#1}}

\newcommand{\Transitions}[1][]{T_{#1}}
\newcommand{\transition}[1][]{t_{#1}}

\newcommand{\FlowRelation}[1][]{F_{#1}}

\newcommand{\Markings}[1][]{M_{#1}}
\newcommand{\marking}[1][]{m_{#1}}
\newcommand{\markingp}[1][]{{\marking}'_{#1}}

\newcommand\prepla[1]{\leftindex{^\bullet{#1}}}
\newcommand\pospla[1]{#1^\bullet}

\newcommand{\lIPN}{\lambda_{pn}}

\newcommand{\HDAName}{HDA\xspace}
\newcommand{\HDAsName}{\HDAName{}s\xspace}
\newcommand{\hda}[1][]{\mathcal{A}_{#1}}
\newcommand{\hdaDef}{\tuple{\SetPrecubical, \Initials}}

\NewDocumentCommand{\Concsets}{o}{
	\IfNoValueTF{#1}
		{\square}
		{\square(#1)}
	}
\NewDocumentCommand{\concset}{O{}}{\tau_{#1}}
\newcommand{\concsetp}[1][]{{\concset}'_{#1}}

\newcommand{\SetPrecubical}[1][]{\mathcal{X}_{#1}}
\newcommand{\SetPrecubicalDef}{\tuple{\Cells, \evHDA, \FaceMaps}}
\newcommand{\setA}[1][]{A_{#1}}
\newcommand{\setB}[1][]{B_{#1}}
\newcommand{\setC}[1][]{C_{#1}}
\newcommand{\setD}[1][]{D_{#1}}
\newcommand{\setE}[1][]{E_{#1}}
\newcommand{\setU}[1][]{U_{#1}}
\NewDocumentCommand{\Cells}{O{}}{X_{#1}}
\newcommand{\CellsType}[1][\setU]{\Cells{}[#1]}
\newcommand{\cell}[1][]{x_{#1}}

\newcommand{\cellZero}[1][]{z_{#1}}
\newcommand{\cellS}[1][S]{\cell[#1]}
\newcommand{\cellT}[1][T]{\cell[#1]}
\newcommand{\evHDA}{\textfun{ev}}
\newcommand{\FaceMaps}[1][]{\Delta_{#1}}

\NewDocumentCommand{\facemap}{O{\setA,\setB;\setU}}{\delta_{#1}}
\NewDocumentCommand{\facemapdown}{O{\setA}}{\delta_{#1}^0}
\NewDocumentCommand{\facemapup}{O{\setB}}{\delta_{#1}^1}

\newcommand{\ktrunc}[2][k]{{#2}^{\leq #1}}

\newcommand{\Initials}[1][]{I_{#1}}

\newcommand{\pathHda}[1][]{\rho_{#1}}
\newcommand{\pathHdaDef}{(\cellZero[0], \cell[0], \cellZero[1], \dotsc, \cellZero[{n-1}], \cell[{n-1}], \cellZero[n])}
\newcommand{\pathHdaInfDef}{(\cellZero[0], \cell[0], \cellZero[1], \dotsc)}

\newcommand{\Computations}{\textfun{Comp}}

\newcommand{\lIHDA}{\lambda_{HDA}}

\NewDocumentCommand{\PNcell}{O{\marking}O{\concset}}{(#1, #2)}
\NewDocumentCommand{\PNcellp}{O{\markingp}O{\concsetp}}{(#1, #2)}
\NewDocumentCommand{\PNcellInitial}{O{}O{}}{\gamma}

\newcommand{\pntohda}[1]{\textfun{pn2HDA}(#1)}

\newcommand{\PNInitialsDef}[1][{\marking[0]}]{\{\PNcell \mid \marking = #1\}}

\NewDocumentCommand{\Graph}{O{}D<>{}}{G_{#1}^{#2}}

\newcommand{\PNGraph}[1][]{\Graph[\petrinet[#1]]<1>}

\newcommand{\tuple}[1]{\langle #1 \rangle}

\newcommand{\union}{\cup}

\newcommand{\Size}[1]{|#1|}

\newcommand{\Naturals}{\mathbb{N}}
\newcommand{\Bools}{\mathbb{B}}

\newcommand{\isomorph}{\simeq}

\newcommand{\Range}[2][0]{\{#1, \dotsc, #2\}}

\newcommand{\nth}[2][]{{#1}[#2]}

\newcommand{\Multiset}[1][]{f_{#1}}

%% file: intro.tex
Modern Cyber-Physical Production Systems (CPPS) are increasingly distributed and concurrent, which makes the design,  by discrete-event systems (DES) formalism, and validation of controllers more challenging \cite{alurPrinciplesCyberPhysicalSystems2015}. While formal DES methods provide strong guarantees for verification, diagnosis, and synthesis, their practical effectiveness depends on how well the chosen semantics reflects what actually happens at execution time \cite{songCyberPhysicalSystemsFoundations2016}. In many industrial settings (PLC scan cycles, asynchronous I/O, communication latencies), actions that are treated as instantaneous in the model may occur effectively simultaneously, or with an order that is not under the designer’s control \cite{cantarelliReactiveControlSystem2008}.

A common modeling choice is to represent concurrency through interleavings: concurrent events are serialized into sequences \cite{cantarelliReactiveControlSystem2008}. This abstraction is often sufficient for reachability-style questions, but it can hide concurrency-induced inconsistencies that only appear when multiple events occur “at the same time.” A typical example is the assignment of outputs: two events that are independent in the net (hence potentially concurrent in reality) may impose conflicting output updates, yet the conflict can remain invisible when those events are forced to occur in some order. Such issues are especially problematic when specifications are partial or loosely constrained, since multiple implementations may satisfy the interleaving model while behaving differently under concurrent executions \cite{rousselAlgebraicSynthesisLogical2012}.

This paper proposes a true-concurrency viewpoint for control specifications by leveraging Higher-Dimensional Automata (HDA) \cite{DBLP:journals/tcs/Glabbeek06}, where higher-dimensional cells explicitly represent concurrent steps rather than encoding them as arbitrary interleavings. We introduce here an interpreted variant, called Interpreted HDA (IHDA) and based on Interpreted Petri Net (IPN) \cite{davidPetriNetsModeling1994}, to include inputs/outputs and make the model directly usable for automation control. In IHDA, concurrency is not only represented, it becomes an analyzable object, allowing for verifications that are formulated at the level of simultaneous actions.

Our main contribution is a translation from Interpreted Petri Nets (IPN) to IHDA that preserves the standard sequential behavior while explicitly generating higher-dimensional cells for admissible concurrent firings. On top of this construction, we propose a simple mechanism to detect output inconsistencies under concurrency: if a concurrent step entails incompatible output requirements (e.g., simultaneous opposite updates), the inconsistency becomes explicit in the corresponding cell constraints, and a witness execution can be extracted. We illustrate the approach on a representative manufacturing example implemented in a simulation tool, showing how IHDA helps bridge the gap between Discrete-Event System controller designed in common formalism (Automaton or Petri Net) and the concurrency effects induced by the physical execution layer.

The remainder of the paper is organized as follows. In section \ref{sec:IPN},  we recall background on Interpreted Petri Net, while recall on HDA is in section \ref{sec:HDA}. The section \ref{sec:IHDA} is dedicated to define IHDA and the IPN to IHDA translation. This section  presents also the inconsistency detection principle. Finally, section \ref{sec:ME} presents a case study of manufacturing system designed in a simulation tool.

%% file: pndef.tex
Interpreted Petri Nets (IPNs) are a well known formalism in the control and automatisation community in order to use 
Petri nets as effective plant controllers. To this end, the Petri net is extended with input and output signals allowing to interface it with the physical plant \cite{davidPetriNetsModeling1994,Saives_example}.

The inputs and outputs of IPNs are often directly associated to a sensor or an actuator.
In this work take a slightly more generic approach. We use sets of atomic propositions (APs) as input and output, 
as it is standard when working with (input-output) automata \cite{DBLP:conf/cav/Duret-LutzRCRAS22, DBLP:conf/atva/KretinskyMS18} or
in the context of reactive synthesis \cite{DBLP:journals/sttt/JacobsPABCCDDDFFKKLMMPR24}. 
An atomic proposition correspond to a Boolean variable and transitions in automata can be labelled by Boolean functions over these
propositions rather than simple letters. While this does not change the expressiveness of the automaton, it greatly increases practical usability.

Here we take a similar approach for defining the inputs and outputs of the IPN.
To keep notions concise, we will use a subset of the Boolean functions called cubes.
A cube is a Boolean function which uses only conjunctions, atomic propositions and their negations (i.e. no disjunction).
They are generated by the grammar $c \Coloneqq ap \mid \neg ap \mid c \land c$ with $c$ being a cube and $ap$ being a atomic proposition.
Finally we call a cube that uses all of the atomic propositions (each of the APs is associates to $\top$ or $\bot$) a valuation.
This restriction, that makes the presentation easier, can however easily be removed in the future.

Formally we have:
\begin{definition}
\label{def:ipn}
    An \emph{Interpreted Petri Net (IPN)} $Q = \left\lbrace G, M_0, I, O, \lIPN \right\rbrace$ where $(G, M_0)$ is a standard marked Petri net with $G = (P, T)$ being the places and transition and 
    \begin{itemize}
        \item $I$ is a finite set of atomic input propositions,
        \item $O$ is a finite set of atomic output propositions,
        \item $\lIPN$ being the labeling function.
    \end{itemize}
    The labeling function $\lIPN$ associates to each place in the original net a cube
    over the outputs and to each transition a pair of cubes, one over the inputs and one over the outputs, denoted $\lIPN(t)[0]$/$\lIPN(t)[1]$ for the input/output condition.
\end{definition}

A standard Petri net in this context is unweighted, uncolored and without inhibitor arcs or any other extension.
This is a deliberate choice in order to keep the definitions compact. A straightforward adaption of many of these
extensions is however possible, as only the construction of the corresponding HDA (see Section~\ref{sec:HDA}) needs to 
be changed and dedicated constructions for many of these techniques are introduced in \cite{DBLP:conf/apn/AmraneBFHS25}.

The firing of transitions in such a Petri net is conditioned by the inputs. That is a transition $t$ can only fire when (as for a conventional Petri net) all preplaces $\prepla{t}$ have at least one token \emph{and} the current valuation of the inputs is compatible with $\lIPN(t)[0]$.
Formally, the input is given as a valuation over the input propositions, denoted $i$, and it is compatible with a transition $t$ iff
$i \models \lIPN(t)[0]$.

An interpreted execution of such a Petri net transforms a stream of input valuations into a stream of output valuations. We interpret the combination of these two streams as infinite words: $w = (i_0, o_0)(i_1, o_1)(i_2, o_2),\dotsc \in (\Bools^{\left|I\right|}, \Bools^{\left|O\right|})^\omega$. We say that such a word $w$ is compatible with the IPN $Q$ if there exists a run $r = m_0m_1m_2m_3\dotsc \in {\Naturals^{\left|P\right|}}^\omega$ 
in $Q$ such that between two markings either a transition is taken or the marking stays the same as no transitions are fireable due to the current marking and/or the current valuation of the inputs. Additionally, the outputs must be compatible.

Formally we have 
\begin{itemize}
    \item Wait Step $k$: $m_k = m_{k+1}$ therefore there is no transition  $t \in T$ such that $m_k \ge \prepla{t}$ and $i_k \models \lIPN(t)[0]$ both hold and we have $o_k \models \bigwedge_{p \in P \colon m_k[p] > 0} \lIPN(p)$
    \item Transition Step $k$: Here a transition $t$ is taken (therefore marking and input constraints must be verified), and we have $o_k \models \bigwedge_{p \in P \colon m_k[p] > 0} \lIPN(p) \land \lIPN(t)[1]$ and the marking evolves according to $m_{k+1} = m_k - \prepla{t} + \pospla{t}$.
\end{itemize}

Intuitively, the output produced corresponds to the conjunction of the place outputs and, in the case of a transition step, the output constraints associated to the fired transition.

%% file: hdadef.tex
A \emph{Higher-Dimensional Automaton} (\HDAName for short) is a mathematical framework that represents the true concurrency of a system. Intuitively, a \HDAName is a set of \emph{cells} labelled with actions of the modelled system that may occur concurrently. These cells are connected through \emph{face maps}: a face map is a function that defines which cells can be reached from  by terminating or unstarting a subset of the currently active actions.
In recent work~\cite{DBLP:conf/apn/AmraneBFHS25}, we present how a \HDAName can be used to model a Petri net, where the aforementioned actions instantiate the transitions of the Petri net. Since firing the same transition of a Petri net twice induces the same change on the current marking and can not be differentiated otherwise\footnote{See \cite{DBLP:journals/tcs/Glabbeek06} for a detailed discussion of individual and collective token semantics. We use the \textit{standard} collective token semantic here.}, we do not need to keep track of the identity of a transition. They all behave the same and are exchangeable. Thus, we can here rely on anonymous \HDAName, which makes the formalism simpler. In the rest of the paper, when referring to \HDAName, we mean anonymous \HDAName.


\subsection{Formal definition}
\label{sec:HDAformaldef}

Given an alphabet $\Alphabet$, we denote by $\Concsets[\Alphabet]$, or $\Concsets$ when the alphabet is clear from the context, the set of multisets over $\Alphabet$. Such a multiset of $\Alphabet$ is called a concurrent set, or \emph{concset} for short.

Recall that multisets are generalizations of sets where elements can be present more than once. Formally, a multiset $\Multiset$ over a set $\setE$ is a function $\Multiset: \setE \to \Naturals$. We use the usual notations $\Multiset[1] + \Multiset[2]$ for the multiset $\Multiset$ such that for every $e \in \setE$, we have $\Multiset(e) = \Multiset[1](e) + \Multiset[2](e)$; we also write $\Multiset[1] \leq \Multiset[2]$ when for every $e \in \Multiset[1](e) \leq \Multiset[2](e)$ and finally, when $\Multiset[1] \leq \Multiset[2]$, we write $\Multiset[2] - \Multiset[1]$ for the multiset $\Multiset$ such that for every $e\in\setE$, we have $\Multiset(e) = \Multiset[2](e) - \Multiset[1](e)$ 

\begin{definition}
\label{def:precubicalset}
    An \emph{(anonymous) precubical set} $\SetPrecubical = \SetPrecubicalDef$ consists of a set of \emph{cells} $\Cells$ together with a function $\evHDA: \Cells \to \Concsets$ and a set of \emph{face maps} $\FaceMaps$.
For a concset $\setU$ we write $\CellsType = \{\cell \in \Cells \mid \evHDA(\cell) = \setU \}$ for the cells of type $\setU$.
Further, for every $\setU, \setA, \setB \in \Concsets$ with $\setA + \setB \leq \setU$ there are face maps $\facemap: \CellsType \to \CellsType[\setU - (\setA + \setB)]$
which satisfy
\begin{equation}
	\label{eq:precid}
	\facemap[{\setC, \setD; \setU - (\setA + \setB)}] \facemap = \facemap[{\setA + \setC, \setB + \setD; \setU}]
\end{equation}
for every $\setU \in \Concsets$, $\setA + \setB \leq \setU$, and $\setC + \setD \leq \setU -(\setA + \setB)$.
\end{definition}


We will omit the extra subscript ``$\setU$'' in the face maps
and further write $\facemapdown = \facemap[\setA, \emptyset]$ and $\facemapup = \facemap[\emptyset, \setB]$.
The \emph{upper} face maps $\facemapup$ transform a cell $\cell$ into one in which the transitions in $\setB$ have been terminated;
the \emph{lower} face maps $\facemapdown$ transform $\cell$ into a cell where the transitions in $\setA$ have not yet started, that is they have been unstarted seen from the $\cell$.
Every face map $\facemap[{\setA, \setB}]$ can be written as a composition $\facemap[{\setA, \setB}] = \facemapdown \facemapup = \facemapup \facemapdown$,
and the \emph{precubical identity} \eqref{eq:precid}
expresses  that these transformations commute.

We write $\Cells[n] = \{\cell \in \Cells \mid \Size{\evHDA(\cell)} = n\}$ for $n \in \Naturals$ and call elements of $\Cells[n]$ \emph{$n$-cells}.
The \emph{dimension} of $\cell \in \Cells$ is $\dim(\cell) = \Size{\evHDA(x)} \in \Naturals$;
the dimension of $\Cells$ is $\dim(\Cells) = \sup\{\dim(\cell) \mid \cell \in \Cells\}\in \Naturals \union\{\infty\}$.
For $k \in \Naturals$, the \emph{$k$-truncation} of $X$
is the precubical set $\ktrunc{\Cells}=\{\cell \in \Cells \mid \dim(\cell) \leq k\}$ with all cells of dimension higher than $k$ removed.

A \emph{higher-dimensional automaton} \emph{(\HDAName)}
$\hda = (\Alphabet, \SetPrecubical, \Initials)$ consists of
a finite alphabet $\Alphabet$,
a precubical set $\SetPrecubical = \SetPrecubicalDef$ on $\Alphabet$, and a subset $\Initials \subseteq \Cells$ of initial cells.
We will not need accepting cells and usually have a single initial cell in this work.
In general a \HDAName may be finite, infinite, or even infinite-dimensional, however we restrict ourselves to finite-dimensional finite \HDAName,
as it is done in \cite{DBLP:conf/apn/AmraneBFHS25} which is sufficient to represent bounded (also called $k$-safe) Petri Nets. We naturally lift all notations for precubical sets to \HDAsName.

For this work, we give the so-called \emph{concurrent step} semantics to \HDAsName when defining computations. A concurrent step computation is a sequence 
\begin{equation}
    \label{eq:def_paths}
    \pathHda = \pathHdaDef
\end{equation}
where $\cellZero[i]$ are $0$-cells and $\facemapdown[\evHDA(\cell[i])](\cell[i])=\cellZero[i]$ for every $i\in\Range{n-1}$ and $\facemapup[\evHDA(\cell[{i-1}])](\cell[i-1]) = \cellZero[i]$ for every $i\in\Range[1]{n}$. 
The set of computations of an \HDAName $\hda$ is denoted by $\Computations(\hda)$.

Intuitively, $\cellZero[i]$ marks the change of control cycle and the $\cell[i]$ correspond to the set of all actions happening between $\cellZero[i]$ and $\cellZero[i+1]$.

A cell $\cellT \in \Cells$ is \emph{reachable} from $\cellS$ if there exists a path $\pathHda$ from $\cellS$ to $\cellT$,
\ie $\cellZero[0] = \cellS$ and $\cell[n] = \cellT$ in the notation (\ref{eq:def_paths}) above.

\subsection{From Petri Nets to \HDAsName}
\label{sec:pn2HDA}

In~\cite{DBLP:conf/apn/AmraneBFHS25}, a Petri net $\petrinet = \petrinetDef$ can be translated into a \HDAName $\pntohda{\petrinet} = \hdaDef$ with $\SetPrecubical = \SetPrecubicalDef$ defined as follows.
Let $\Concsets = \Concsets[\Transitions]$ and define $\Cells = \Markings \times \Concsets$ and $\evHDA: \Cells \to \Concsets$ by $\evHDA(\PNcell) = \concset$ with $\Markings$ being the markings of the Petri net in $\Naturals^{\left|P\right|}$.
For $\cell = \PNcell \in \Cells[\concset]$ with $\concset = (\transition[1], \dotsc, \transition[n])$ non-empty and $i \in \Range[1]{n}$, define $\FaceMaps$ by
\begin{align*}
	\facemapdown[\transition[i]](\PNcell) &= (\marking + \prepla{\transition[i]}, (\concset - \transition[i])), \\
	\facemapup[\transition[i]](\PNcell) &= (\marking + \pospla{\transition[i]}, (\concset - \transition[i])).
\end{align*}

Intuitively, the $0$-cells and the $1$-cells of $\hda$ respectively correspond to the vertices and the edges of the reachability graph of $\petrinet$. Cells of higher dimension represent multiple transitions fired concurrently.

\begin{lemma}[4 in \cite{DBLP:conf/apn/AmraneBFHS25}]
    \label{lmm:pnHDAequiv}
    The reachability graph of Petri net $\petrinet$ is isomorphic to the $1$-truncation of $\pntohda{\petrinet}$: $\PNGraph \isomorph \ktrunc[1]{\pntohda{\petrinet}}$
\end{lemma}

The original paper provides a translation to non-anonymous \HDAName, which is a more general setting allowing the differentiation of two concurrent actions with the same label. Formally, instead of concset, non-anonymous \HDAsName rely on conclist, which are ordered sequences of events labeled with actions. This more general setting is not necessary when working with Petri nets and was resolved in \cite{DBLP:conf/apn/AmraneBFHS25}
by imposing an arbitrary (lexicographic) order an the actions. Here we use the more appropriate formalism
of anonymous HDA established in \cite{KrzysztofPrivate}.


%% file: ihdadefPSC.tex
\subsection{Definitions}

\begin{definition}
\label{def:ihda}
    An \emph{Interpreted anonymous HDA (IHDA)} is defined by the four-tuple $\hda[I] = \left\lbrace \hda, I, O, \lIHDA \right\rbrace$ where $\hda$ is a standard HDA 
    but with cells being conditioned over the inputs and associated to an output. 
    Formally we have
    \begin{itemize}
        \item $I$ is a finite set of atomic input propositions,
        \item $O$ is a finite set of atomic output propositions,
        \item $\lIHDA$ being the labeling function.
    \end{itemize}
    The labeling function $\lIHDA$ associates to each cell $\cell$ in the $\hda$ a pair of cubes, one over the inputs $\lIHDA(\cell)[0]$ and one of the outputs $\lIHDA(\cell)[1]$.
\end{definition}

Interpreted executions of IHDAs show an input/output behavior, much like IPNs, producing words of the form $w = (i_0, o_0)(i_1, o_1)(i_2, o_2),\dotsc \in (\Bools^{\left|I\right|}, \Bools^{\left|O\right|})^\omega$. We say that such a word $w$ is compatible with the concurrent step semantic of IHDA $\hda[I]$ if there exists a computation $\pathHda = \pathHdaInfDef$ that is compatible in the following sense:

We use two counters $n_0=0$ and $n_1=0$ to track the current position in $w$ and $\pathHda$.
At all times we need to have $i_{n_0} \models \lIHDA(\nth[\pathHda]{n_1})[0]$ and $o_{n_0} \models \lIHDA(\nth[\pathHda]{n_1})[1]$. That is the current cell is compatible with the inputs and produces compatible outputs.
Finally, when reading the word $w$ by incrementing the $n_0$ counter, we can use one of two rules:

\begin{itemize}
    \item Only $n_1$ is incremented (Akin to a ``wait step''),
    \item $n_0$ and $n_1$ are incremented (Akin to a ``transition step'').
\end{itemize}

\subsection{From IPN to IHDA}
\label{subsec:ipn2ihda}

The main interest of IHDA is to represent the possibility of concurrent step semantics in a canonical fashion.
We will therefore discuss the translation of an IPN into a corresponding IHDA and how, on one hand, this helps to increase expressiveness and on the other hand automatically identifies inconsistencies in the output assignment.

The first step of the construction employs the technique recapitulated in section~\ref{sec:pn2HDA} to translate a standard Petri Net into an HDA,
with the additional constraint that only input compatible cells, that is for $\cell = (m,c)$ we have $\bigwedge_{t \in c}\lIPN(t)[0] \not\equiv \bot$, are constructed. Note that cells with dimension of $0$ or $1$ are always input compatible and therefore the reachability graph of the PN underlying the IPN is isomorphic to the 1-truncation of the HDA underlying the IHDA. The second step is to derive the labeling function of the HDA from the labelling function of the IPN in a consistent manner. Given a cell $\cell = (m, c)$ defined by its marking and concset, we define the labeling function
\begin{multline*}
    \lIHDA(\cell) = \\ 
    \left(\bigwedge_{t \in c}\lIPN(t)[0], \bigwedge_{p\in P : m[p] > 0 \lor p \in \prepla{c}}\lIPN(p) \land \bigwedge_{t \in c}\lIPN(t)[1] \right),
\end{multline*}
where $\prepla{c}$ extends preplaces to concsets: $\prepla{c} = \sum_{t \in c} c(t)\times\prepla{t}$.

As we only construct input compatible cells we have $\lIHDA(\cell)[0] \not\equiv \bot$ for all cells. 
With these definitions set, we can discuss the first point: detection of output inconsistencies. 
The concsets of the cells correspond to the input compatible concurrently fireable transitions of the IPN.
That is there exists a reachable marking $m$ and an input valuation such that $c$ is a valid concstep,
that is $m \ge \prepla{c}$ and $\bigwedge_{t \in c}\lIPN(t)[0] \not\equiv \bot$.
$\lIHDA(\cell)[1]$ represents the output generated by the cells. If there exists a cell such that $\lIHDA(\cell)[1] \equiv \bot$, this means that the different output conditions contradict one another. Such a contradiction, like demanding our chariot going left and right at the same time, correspond to a fault in the control architecture and needs to be corrected. The IHDA identifies such situations in a principled manner and can provide traces leading to them.


\subsection{From IHDAs to Controllers}

Before introducing the manufacturing example, we want to give a high-level picture of the technique
proposed and how to get from an IPN to an actual controller.

\begin{figure}[htbp]
\begin{minipage}{0.4\linewidth}
\centering
\begin{tikzpicture}[node distance=.5cm, auto]
    \node[draw, rectangle] (IPN) {IPN};
    \node[draw, rectangle, below=of IPN] (IHDA) {IHDA};
    \node[draw, rectangle, below=of IHDA] (CSC) {CSC};
    \node[draw, ellipse, below=of CSC] (Plant) {Plant};
    
    \node[draw, color=green, rectangle, fit=(IPN) (IHDA) (CSC), inner sep=12pt, thick] (box) {};
    \node[draw, color=red, circle, fit= (CSC) (Plant), inner sep=3pt, thick] (circ) {};

    \draw[->] (IPN) -- (IHDA);
    \draw[->] (IHDA) -- (CSC);
    
    \draw[->] (CSC) to[bend right] node[left] {$o$} (Plant);
    \draw[->] (Plant) to[bend right] node[right] {$i$} (CSC);
\end{tikzpicture}
\end{minipage}
\hfill
\begin{minipage}{0.5\linewidth}
    \captionof{figure}{Pipeline depicting the generation and use of the concurrent step controller (CSC).}
    \label{fig:IPN2CTRL}
\end{minipage}
\end{figure}
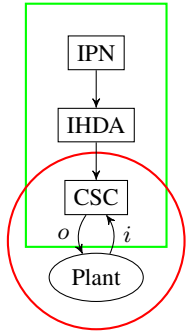

Figure~\ref{fig:IPN2CTRL} shows the pipeline of the proposed approach.
The green box depicts the preprocessing or generation steps. We go from an IPN to an IHDA. 
We then endow it with the concurrent step semantics to generate the concurrent step controller. 
The red circle encloses the physical/simulated plant and the controller. The arrows mark the control cycle. 
In each cycle the plant sends the current valuation of the inputs ($i$) to the CSC which 
in turn responds with a valuation for the outputs ($o$).

%% file: example.tex
As states as contribution of this paper, we present in this section a small example, inspired from \cite{Saives_example}, illustrating the use of HDA to control a manufacturing system.

\subsection{Example presentation}
\begin{figure}[!t]
\centering
\includegraphics[width=0.5\textwidth]{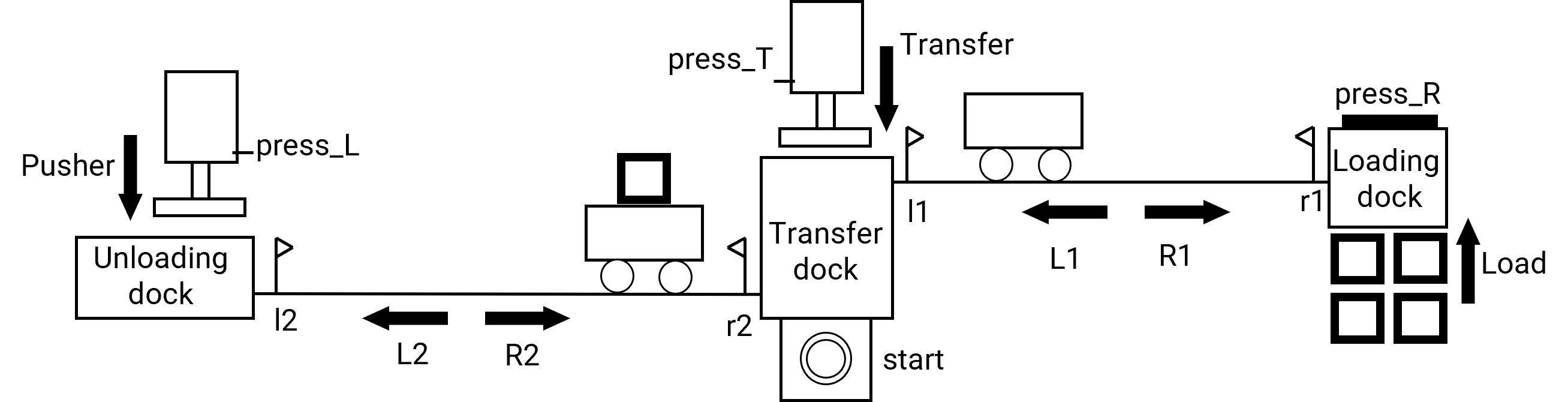}
\caption{Manufacturing example}
\label{manuf_ex}
\end{figure}

The considered system is illustrated on the Fig.\ref{manuf_ex}. In this system, the transfer dock is the central piece, with the \textit{start} button and the initial position of two transport chariots (sensors \textit{l1} and \textit{r2}). The upper chariot can roll right (actuator \textbf{R1}) to reach the loading dock, sensor \textit{r1} indicating its presence. Once charged with a load (the chariot is equipped with a pressure sensor \textit{press\_R}), the chariot goes left (\textbf{L1}) back to transfer dock. In parallel, the lower chariot transports a load to the left (\textbf{L2}) toward the unloading dock. When it reaches position \textit{l2}, a cylinder actuated by \textbf{Pusher} ejects the load. The load release is confirmed by the pressure sensor \textit{press\_L}, while the chariot returns to the transfer dock by moving right (\textbf{R2}).
The transfer of the load between two chariots is operated by the action \textbf{Transfer} and this is done when the sensor \textit{press\_T} is activated. 

%

\begin{figure}[!t]
\centering
\includegraphics[width=0.5\textwidth]{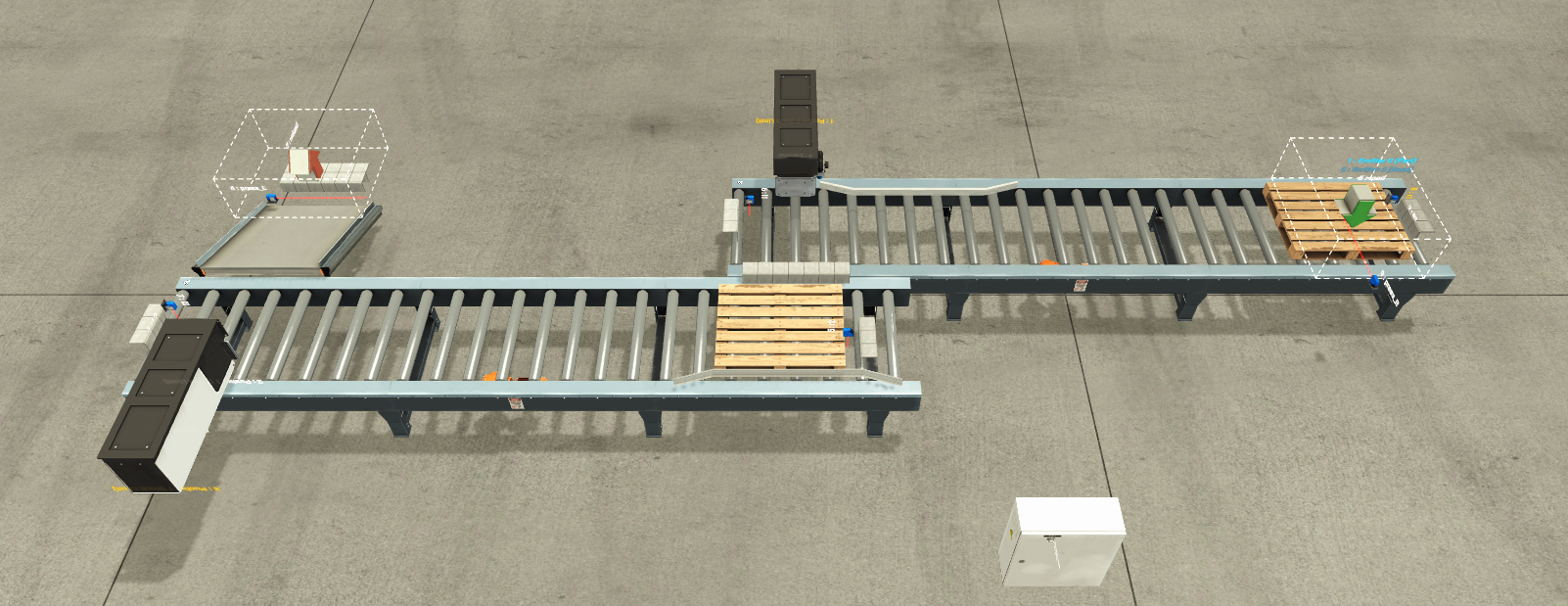}
\caption{Manufacturing example designed in the simulation tool Factory I/O}
\label{FactoryIO}
\end{figure}

In order to simulate this system, we implement it in the simulation tool "Factory I/O" (\cite{FactoryNextGenPLC,rieraHOMEFACTORYVirtual2017}). In the Fig.\ref{FactoryIO}, is shown the designed system in this tool. The interconnection with this system is operated by Modbus TCP and OPC Client DA/UA. This allows to design the control law in Python and connect the inputs/outputs by Modbus TCP and MQTT. 

\begin{figure}[!t]
\centering
\includegraphics[width=0.45\textwidth]{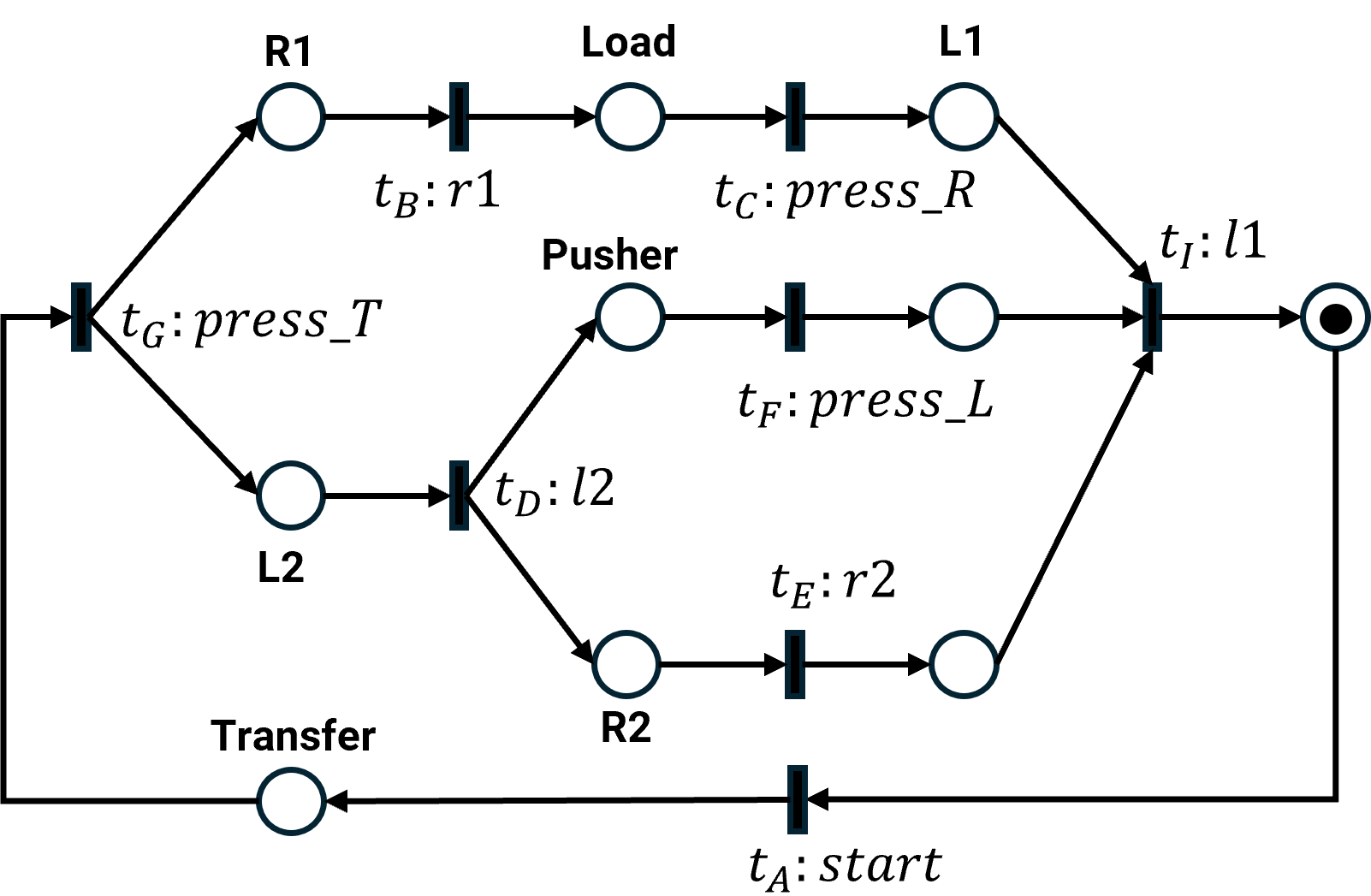}
\caption{Control law designed by Interpreted Petri Net of the manufacturing system}
\label{PN_ex}
\end{figure}

The control law of this system is designed via a Interpreted Petri Net (IPN) and is illustrated on Fig.\ref{PN_ex}. We keep this control law intentionally simple to improve the understanding of the use and implementation of the IHDA.

\subsection{Translation into IHDA}
\begin{figure}
    \centering
    
\begin{tikzpicture}	
		
		\path[fill=black!20!white] (2,0) -- (4,0) -- (4,2) -- (2,2) -- (2,0);
		\node (cell1) at (3,1) {$x$};
		
		\path[fill=black!20!white] (2,2) -- (4,2) -- (4,4) -- (2,4) -- (2,2);
		\node (cell2) at (3,3) {$y$};
		
		\path[fill=black!35!white] (4,0) -- (6,0) -- (7,1) -- (7,3) -- (5,3) -- (4,2) -- (4,0);
		\node (cell3) at (5.5,1.5) {$w$};
		
		\path[fill=black!35!white] (4,2) -- (6,2) -- (7,3) -- (7,5) -- (5,5) -- (4,4) -- (4,2);
		\node (cell4) at (5.5,3.5) {$z$};
		
		\node[state, initial below] (start) at (0,0) {};
		\node[state] (0) at (1,0) {};
		\node[state] (1) at (2,0) {};
		\node[state] (01) at (4,0) {};
		\node[state] (020) at (6,0) {};
		\node[state] (011) at (5,1) {};
		\node[state] (021) at (7,1) {};
		\node[state] (10) at (2,2) {};
		\node[state] (11) at (4,2) {};
		\node[state] (120) at (6,2) {};
		\node[state] (111) at (5,3) {};
		\node[state] (121) at (7,3) {};
		\node[state] (20) at (2,4) {};
		\node[state] (21) at (4,4) {};
		\node[state] (220) at (6,4) {};
		\node[state] (211) at (5,5) {};
		\node[state] (221) at (7,5) {};
		\node[state] (end) at (8,5) {};
		\node (next) at (9,5) {};
		
		\path (start) edge node[swap] {$t_A$} (0);
		\path (0) edge node[swap] {$t_G$} (1);
		
		\path (1) edge node[swap] {$t_D$} (01);
		\path (01) edge node[swap] {$t_F$} (020);
		\path (020) edge node[swap] {$t_E$} (021);
		\path (01) edge (011); 
		\path (011) edge (021);
		
		\path (10) edge (11);
		\path (11) edge (120);
		\path (120) edge (121);
		\path (11) edge (111); 
		\path (111)  edge(121);
		
		\path (20) edge (21);
		\path (21)  edge (220);
		\path (220) edge (221); 
		\path (21)  edge(211);
		\path (211)  edge(221);
		
		\path (1) edge node {$t_B$} (10); 
		\path (10)  edge node {$t_C$} (20);
		
		\path (01) edge (11); 
		\path (11)  edge (21);
		
		\path (020) edge (120); 
		\path (120)  edge (220);
		
		\path (021) edge (121); 
		\path (121)  edge (221);
		
		\path (011) edge (111); 
		\path (111)  edge (211);
		
		\path (221) edge node[swap] {$t_I$} (end);
		\path (end) edge[dotted] node[swap] {$t_A$} (next);
\end{tikzpicture}
    \caption{The HDA corresponding to the Interpreted Petri Net of Figure~\ref{PN_ex}. }
    \label{fig:ihda}
\end{figure}
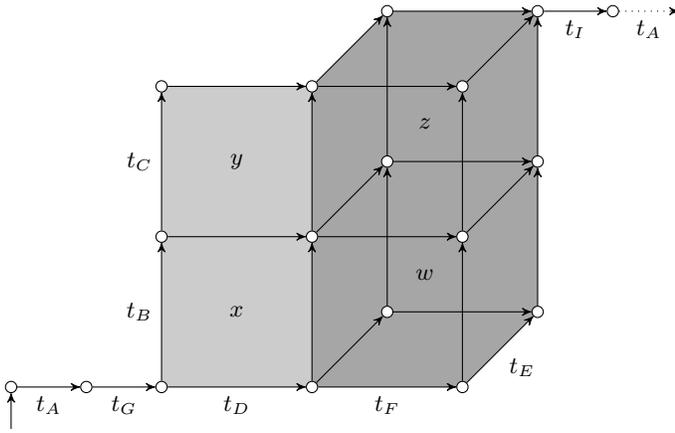

As described in above, we first translate the PN underlying the IPN into a HDA (Section~\ref{sec:HDA}).
Then, the additional information about inputs and outputs is added to convert the HDA into an IHDA 
corresponding to the IPN (Section~\ref{sec:IHDA}).
The so produced IHDA is illustrated on the Fig.\ref{fig:ihda} with the inputs and outputs being specified
in the Table~\ref{tab:IHDA_IO}. To give an example, the $2$-cell $x$ corresponds to the two chariots arriving respectively at the loading and the unloading docks at the same moment and therefore $t_B$ and $t_D$ are taken at the same time; and the $2$-cell 
$y$ corresponds to the upper chariot charged with a new load while the lower chariot arrives at the unloading dock dock. The $3$-cell $w$ corresponds to the upper chariot arriving at loading dock while the lower one is back at the initial position and the Pusher pushes the load to the unloading dock. Finally, the $3$-cell $z$ corresponds to the two chariots arriving back to their starting position while the Pusher pushes the load. The outputs of the $0$-cells is given by the corresponding marking of the IPN; Corresponding to the conjunction of all output conditions of
places with at least one token. For $1$-cells, the input is the one of the transition and the output is given by the marking previous to the transition. For the higher dimensional cells, their inputs/outputs are the following (we omit the subcells of $w$ and $z$ to preserve space and readability):
\begin{center}
\begin{tabular}{c|c|c}
    \label{tab:IHDA_IO}
    Cell & Inputs & Outputs\\
    \hline
    $x$ & $\textit{r1} \land \textit{l2}$ & $\textbf{R1} \land \textbf{L2}$ \\
    $y$ & $\textit{press\_R} \land \textit{l2}$ & $\textbf{Load} \land \textbf{L2}$  \\
    $w$ & $\textit{r1} \land \textit{press\_L} \land \textit{r2}$ & $\textbf{R1} \land \textbf{Pusher} \land \textbf{R2}$ \\
    $z$ & $\textit{press\_R} \land \textit{press\_L} \land \textit{r2}$ & $\textbf{Load} \land \textbf{Pusher} \land \textbf{R2}$
\end{tabular}
\end{center}

%% file: error_detection.tex
Continuing the example of the previous section, we can illustrate how error detection works. Let us assume that the lower chariot may not move while the Pusher is being active.
That means \textbf{L2} / \textbf{R2} and \textbf{Pusher} cannot be performed at the same time. 
We can add this restriction directly to our IPN by changing the output labels of the corresponding places from \textbf{L2} to $\textbf{L2} \land \neg \textbf{Pusher}$, \textbf{R2} to $\textbf{R2} \land \neg \textbf{Pusher}$ and from \textbf{Pusher} to $\neg \textbf{L2} \land \textbf{R2} \land \textbf{Pusher}$.

Another approach is to define global invariants over the outputs, here $\neg \textbf{L2} \lor \neg \textbf{Pusher}$ and $\neg \textbf{R2} \lor \neg \textbf{Pusher}$. Then, when computing the labels of each cell, we can check whether the invariant is satisfied and raise an error if it is not.
This is approach is also easily extensible to cover more complex invariants involving input and output propositions.

In both cases, we can verify that both cells~$w$ and~$z$ (and their sub-cells) do not comply with the given restriction: The output for $w$ for instance becomes $\textbf{R1} \land \textbf{Pusher} \land \textbf{R2} \land \neg\textbf{Pusher}$. This is obviously not possible, the Pusher can not be activated and deactivated at the same time and an according error is raised during the construction.
Note that the cells $x$ and $y$ are not affected by this change. The output formula for $x$
for instance becomes $\textbf{R1} \land \textbf{L2}\land \neg \textbf{Pusher}$, which poses
no problem.

\begin{figure}[h]
\centering
\includegraphics[width=0.45\textwidth]{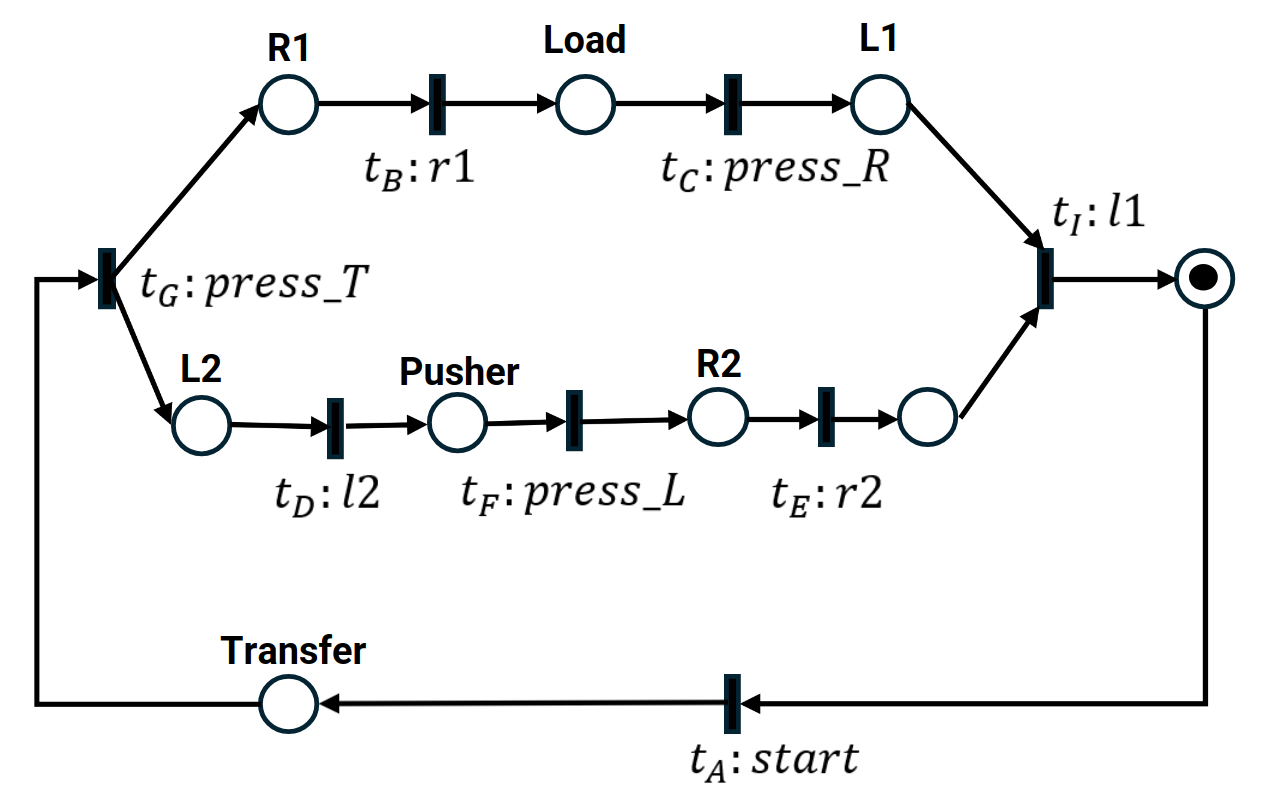}
\caption{Corrected control law designed by IPN for the manufacturing system}
\label{PN_conc}
\end{figure}

\begin{figure}
    \centering
    
\begin{tikzpicture}	
		
		\path[fill=black!20!white] (2*\tikzw,0*\tikzw) -- (4*\tikzw,0*\tikzw) -- (4*\tikzw,2*\tikzw) -- (2*\tikzw,2*\tikzw) -- (2*\tikzw,0*\tikzw);
		\node (cell1) at (3*\tikzw,1*\tikzw) {$x$};
		
		\path[fill=black!20!white] (2*\tikzw,2*\tikzw) -- (4*\tikzw,2*\tikzw) -- (4*\tikzw,4*\tikzw) -- (2*\tikzw,4*\tikzw) -- (2*\tikzw,2*\tikzw);
		\node (cell2) at (3*\tikzw,3*\tikzw) {$y$};
		
		\path[fill=black!20!white] (4*\tikzw,0*\tikzw) -- (6*\tikzw,0*\tikzw) -- (6*\tikzw,2*\tikzw) -- (4*\tikzw,2*\tikzw) -- (4*\tikzw,0*\tikzw);
		\node (cell1) at (5*\tikzw,1*\tikzw) {$z'$};
		
		\path[fill=black!20!white] (4*\tikzw,2*\tikzw) -- (6*\tikzw,2*\tikzw) -- (6*\tikzw,4*\tikzw) -- (4*\tikzw,4*\tikzw) -- (4*\tikzw,2*\tikzw);
		\node (cell2) at (5*\tikzw,3*\tikzw) {$w'$};
		
		\path[fill=black!20!white] (6*\tikzw,0*\tikzw) -- (8*\tikzw,0*\tikzw) -- (8*\tikzw,2*\tikzw) -- (6*\tikzw,2*\tikzw) -- (6*\tikzw,0*\tikzw);
		\node (cell1) at (7*\tikzw,1*\tikzw) {$z''$};
		
		\path[fill=black!20!white] (6*\tikzw,2*\tikzw) -- (8*\tikzw,2*\tikzw) -- (8*\tikzw,4*\tikzw) -- (6*\tikzw,4*\tikzw) -- (6*\tikzw,2*\tikzw);
		\node (cell2) at (7*\tikzw,3*\tikzw) {$w''$};
		
		\node[state, initial below] (start) at (0,0) {};
		\node[state] (0) at (1*\tikzw,0) {};
		\node[state] (00) at (2*\tikzw,0) {};
		\node[state] (01) at (2*\tikzw,2*\tikzw) {};
		\node[state] (02) at (2*\tikzw,4*\tikzw) {};
		\node[state] (10) at (4*\tikzw,0) {};
		\node[state] (11) at (4*\tikzw,2*\tikzw) {};
		\node[state] (12) at (4*\tikzw,4*\tikzw) {};
		\node[state] (20) at (6*\tikzw,0) {};
		\node[state] (21) at (6*\tikzw,2*\tikzw) {};
		\node[state] (22) at (6*\tikzw,4*\tikzw) {};
		\node[state] (30) at (8*\tikzw,0) {};
		\node[state] (31) at (8*\tikzw,2*\tikzw) {};
		\node[state] (32) at (8*\tikzw,4*\tikzw) {};
		
		\node[state] (end) at (9*\tikzw,4*\tikzw) {};
		\node (next) at (10*\tikzw,4*\tikzw) {};
		
		\path (start) edge node[swap] {$t_A$} (0);
		\path (0) edge node[swap] {$t_G$} (00);
		
		\path (00) edge node {$t_B$} (01);
		\path (01) edge node {$t_C$} (02);
		
		\path (00) edge node[swap] {$t_D$} (10);
		\path (01) edge (11); 
		\path (02) edge (12);
		\path (10) edge (11);
		\path (11) edge (12);
		
		\path (10) edge node[swap] {$t_F$} (20);
		\path (11) edge (21); 
		\path (12) edge (22);
		\path (20) edge (21);
		\path (21) edge (22);
		
		\path (20) edge node[swap] {$t_E$} (30);
		\path (21) edge (31); 
		\path (22) edge (32);
		\path (30) edge (31);
		\path (31) edge (32);
		
		\path (32) edge node[swap] {$t_I$} (end);
		\path (end) edge[dotted] node[swap] {$t_A$} (next);
\end{tikzpicture}
    \caption{The HDA corresponding to the Interpreted Petri Net of Figure~\ref{PN_conc}. }
    \label{fig:IHDACORR}
\end{figure}
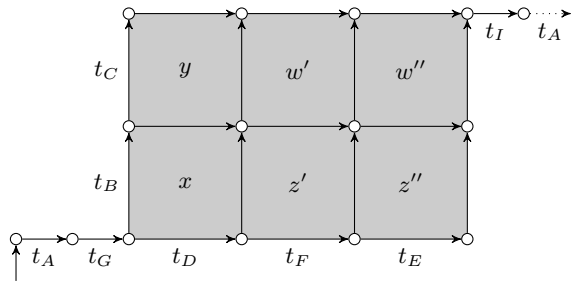

From a run starting at the initial cell and reaching $w$ or $z$, we can derive an execution of the
system in which the Pusher is activated at the same time as the lower chariot is moved to the right. 
There are multiple ways to ensure that pusher has terminated before the chariot starts to 
move again.
Here we have simply put them in ``sequence'', avoiding the branching.
Note that adding a mutual exclusion place between $t_E$ and $t_F$ is not sufficient.
We need to prevent the places causing \textbf{R2} and \textbf{Pusher} to have a token
at the same time.
The corrected IPN is shown in Figure~\ref{PN_conc}, the corresponding IHDA is shown in 
Figure~\ref{fig:IHDACORR}.

Intuitively speaking, each of the 3-cells $w$ and $z$ in the original HDA got decomposed into two 2-cells
($w'$/$w''$ and $z'$/$z''$ respectively) disallowing the concurrent activation of \textbf{Pusher}
and \textbf{R2}.

%% file: simulation.tex

We implemented the IHDA, illustrated by the Fig.\ref{fig:ihda} as a controller interfaced with the manufacturing system modeled in the simulation tool. The resulting closed-loop behavior as well as additional information about the input and output format and how to use 
the IPN to controller translation can be found at \url{https://philippschlehubercaissier.github.io/pages/wodes26.html}. 

The video illustrates the system operating with a 1-second sampling period (inputs and outputs are updated every second). After initialization, the system waits for the \textit{start} action, performs one complete cycle, and finally returns to its initial configuration, i.e., with a box waiting at the transfer dock.

%% file: conclusion.tex
In this paper, we presented the mathematical framework of (anonymous) Interpreted Higher Dimensionnal Automata as a solution for modelling discret-event systems with concurrency. We provided a translation from Interpreted Petri Nets to IHDAs that preserves all the possible behaviors described by the IPN with the addition of the concurrency allowed by the new framework. We then provided an example to instantiate this translation and showcase how it can be used to detect errors in the design of the system. The example has been interfaced and tested with the simulation tool Factory I/O.

This work is a first step in the direction of using the HDA framework for industrial problems. 
We deliberately constrained ourselves to the concurrent step semantics of HDAs but there are other more powerful semantics that could help with other aspects of concurrency in concurrent systems. Notably, HDAs allow for reasoning with non-atomic events - actions are no longer necessarily discrete but may take time.
This allows for a more fine-grained analysis of how actions start, terminate and run concurrently and therefore also a more detailed view of the controlled process.